# Horcrux: A Password Manager for Paranoids

Hannah Li, David Evans
University of Virginia
[hannahli,evans]@virginia.edu

## ABSTRACT

Vulnerabilities in password managers are unremitting because current designs provide large attack surfaces, both at the client and server. We describe and evaluate Horcrux, a password manager that is designed holistically to minimize and decentralize trust, while retaining the usability of a traditional password manager. The prototype Horcrux client, implemented as a Firefox add-on, is split into two components, with code that has access to the user's master's password and any key material isolated into a small auditable component, separate from the complexity of managing the user interface. Instead of exposing actual credentials to the DOM, a dummy username and password are autofilled by the untrusted component. The trusted component intercepts and modifies POST requests before they are encrypted and sent over the network. To avoid trusting a centralized store, stored credentials are secret-shared over multiple servers. To provide domain and username privacy, while maintaining resilience to off-line attacks on a compromised password store, we incorporate cuckoo hashing in a way that ensures an attacker cannot determine if a guessed master password is correct. Our approach only works for websites that do not manipulate entered credentials in the browser client, so we conducted a large-scale experiment that found the technique appears to be compatible with over 98% of tested login forms.

## 1 INTRODUCTION

Users are frequently beseeched to come up with unique, strong passwords for every on-line account, but remembering more than a few high-entropy passwords is well beyond the capabilities of normal humans. Das reported that 43–51% of Internet users reuse the same password across multiple sites [12]. According to a recent survey of security experts, using a password manager was among the most widely-accepted recommendations for improving security [24].

Current password managers, however, do not provide adequate protection for paranoid users. Vulnerabilities are frequently discovered in both the client and server components, as revealed in several recent reports of compromises to major commercial password managers [23, 27, 47, 48]. Password managers which store user credentials in cloud databases have been susceptible to theft and successful dictionary attacks on stolen encrypted data, according to recent news reports on LastPass [49] and KeePass [23]. Password manager critics often cite database theft as a compelling reason not to use password managers: "when password managers fail, they offer a one-stop destination for hackers to obtain all of a target's passwords" [23].

**Contributions.** We present the comprehensive design and evaluation of a password manager, Horcrux,[1] that provides a level of security and privacy well beyond what is achieved by current systems. Horcrux is designed to minimize exposure of secrets to a small, auditable component. Although our design is intended for direct integration into a browser (see Section 7), we have demonstrated and evaluated its effectiveness by implementing a prototype as an open source Firefox add-on (Section 5).

Our server side design (Section 3.2) distributes trust by using secret sharing to store passwords across multiple hosts, and makes novel use of cuckoo hashing to provide user and domain privacy without enabling off-line attacks on password stores. Our client design isolates the component that has access to passwords from the rest of the password manager. Horcrux never exposes the credentials to the DOM, minimizing exposure of user credentials by replacing autofilled dummy credentials with real credentials in outgoing network traffic (Section 3.3). Ideas similar to the password swapping in intercepted outgoing network traffic we use have been proposed before [46], but not adopted by password managers due to usability and compatibility concerns. To evaluate the deployability of our design, we conducted a large-scale study of Alexa's top million websites. As reported in Section 6, we find that Horcrux appears to be compatible with 98% of the websites where a login form was found.

Security researchers have long advocated for minimizing trusted computing bases [13, 30, 42] and using privilege separation [8, 41], and secret-sharing is a well established technique [43]. The main contribution of this work is combining established techniques and our privacy-preserving credential storing scheme in a holistic way to solve an important security problem, and performing a comprehensive evaluation of both the security and compatibility of that design.

## 2 CONTEXT

This section provides background on vulnerabilities in current password manager designs and implementations, and presents the threat model that drives our design decisions.

### 2.1 Password Manager Vulnerabilities

We divide password manager vulnerabilities into client-side vulnerabilities, where the adversary is able to compromise the client or its network connection, and server-side vulnerabilities, where the adversary compromises the password store.

---

[1] A "Horcrux" is a dark, magical object in the *Harry Potter* book series in which a witch or wizard may hide a fragment of their soul. The antagonist of the series, Voldemort, uses multiple horcruxes to distribute the trust of his soul into multiple objects. Voldemort cannot die as long as at least one of his horcruxes is alive.

**Client Vulnerabilities.** Major password managers autofill user credentials into authentication forms found on visited webpages. This exposes those credentials to the DOM, where they are visible to injected malicious scripts [29, 45, 46] that can read the form using `Element.value`. This returns the string currently present in an input field, even if it is a password type field that displays as asterisks. Any value filled in by the password manager or typed in by the user can be stolen by the script. The dynamic nature of JavaScript means functions used in the password manager code may be replaced by adversaries. Li et al. found that all three of the tested password managers that supported bookmarklets were vulnerable to these kinds of vulnerabilities [29]. Silver et al. examined the autofill policies of 15 password managers and found that all had followed some unsafe practices with the security of their autofill policies [45]. None of the password managers comprehensively checked whether the protocol and action of the login form are secure before autofilling the credentials, making the credentials susceptible to being sent to a different destination or through an insecure protocol (HTTP). This risk may be exacerbated by autofill policies that autofill without any user interaction required, allowing attackers to obtain credentials from a large number of domains through an access point.

Password manager clients are also vulnerable because of implementation bugs, particularly in complex URL parsing code. In 2016, Karlsson [26] detected a bug in the Lastpass client script code that treated the wrong part of the URL as the domain, which allowed an attacker to fool the LastPass browser extension into providing a user's credentials for any stored domain. This motivates our design to separate the complex UI code from the small trusted core.

**Server Vulnerabilities.** Password managers can store credentials either in the cloud or locally on the user's computer. Using local storage eliminates the need to trust an external provider, but means that it is now up to the user (who lacks the physical and technical resources of a cloud provider with a data center) to protect the store, and that there is no way to share credentials across multiple devices. It also means that an attacker who can compromise the user's device would now have access to both the password client and store. Although our design allows for the storage devices to include a mix of user-hosted or local stores, for nearly all users, we expect it is a better option to outsource storage to cloud providers.

Users who store their credentials in the cloud or user-hosted platforms inherit the risk of server-side compromises. Unlike a locally stored password database, an adversary could mount an attack on such servers from anywhere and attempt to steal their data, as happened to LastPass in 2015 [22] and OneLogin in 2017 [44]. These thefts typically expose encrypted versions of users passwords, URLs, and emails for all of their online accounts, which are open to thieves to attempt to decrypt offline. Notably, the LastPass server compromise included leakage of a cryptographic hash of many users masters passwords, which was used to decrypt their sensitive online account login information. Modern password managers seek to mitigate offline attacks by storing encrypted passwords using intentionally slow key derivation functions that amplify the cost of a dictionary attack. Currently, in 2017, it is much more common for password managers to never store the user's master password on any device, and instead rely on the key derivations from the user's master password and other local secrets to decrypt passwords on the servers.

## 2.2 Threat Model

Our focus is on mitigating the risks posed by motivated and capable adversaries who can inject scripts into web pages, compromise the client's network access, and may be able to compromise a server database to acquire a full copy of its contents.

**Client Side.** We assume the adversary can inject scripts into any visited webpage, including pages delivered using HTTPS, but cannot access the server-side code to create vulnerabilities on page. This covers the possibility of script injection vulnerabilities in the trusted website, as well as wireless access point attacks, both of which we consider realistic threats.

We distinguish two levels of XSS attackers: a standard attacker who only has the ability to inject scripts into the page with no knowledge of any other vulnerabilities on the server, and a stronger attacker who has specific knowledge of vulnerable pages hosted on the host server and thus can carry out reflect-aided XSS attacks. The severity of these reflect-aided attacks depends on the vulnerabilities on the page—from a vulnerable login target page, to one that reflects all parameters or certain parameters. The compromised page on the site would then resend the submitted password to another site owned by the attacker, or back to the compromised client where they would now be vulnerable to the attacker's scripts. For now, we consider stronger client-side attackers with knowledge about server-side vulnerabilities out of scope; although some strategies may mitigate certain types of these attacks (for example, restricting the specific target pages to the original target and limiting the names of fields containing password), an attacker with access to both a XSS vulnerability and knowledge of a vulnerable server page has many ways to victimize the user.

In other client-side attacks, an adversary who can compromise the client's browser to access its internal state (for example, by exploiting a memory corruption vulnerability in the browser to executed arbitrary code with a ROP attack) can extract secrets from anywhere in the browser's memory, and can generate authentic-looking dialog boxes to request the master password from the user. Similarly, an attacker with a root-level compromise of the client's system, can install a keylogger or alter the clients certificate store to spoof HTTPS connections to targeted sites. No password manager design can provide strong defenses to compromises at those levels. Hence, we consider the web browser and host operating system to be a trusted computing base.

We do not consider social engineering or interface spoofing attacks where victims would be tricked into entering their master password into a rogue dialog box or directly providing their unencrypted credentials to an adversary's site. The lack of trusted input paths in commonly-used computing systems is an important problem and serious threat, but outside the scope of this work. As discussed further in Section 4.1, our prototype implementation also assumes the user will not be tricked into installing a malicious add-on that can observe network traffic after the password manager add-on has inserted the real password (this is necessary for our prototype because of limitations in Firefox's extension mechanism, but would not be an issue for a password manager built into a browser or mobile OS).

**Server Side.** For the server side, we include the threat of full release of all data stored by a cloud server. This could happen as the result of a server vulnerability [19], insider attack from a cloud service employee, or the cloud service complying with a subpoena or national security letter. Hence, it is important that the data stored by the cloud servers is not vulnerable to an offline guessing attack on the master password. This leads to a design objective that everything encrypted with keys derived from the master password must be indistinguishable to an attacker so there is no way to determine if a guess is correct.

Finally, we assume the user should not have any intrinsic trust in the provider of the password manager. This means we want a design where the code that has access to sensitive data is as small and simple as possible, to make individual or third-party auditing realistic.

## 3 DESIGN

Our prototype password manager, Horcrux, is implemented as an open-source Firefox add-on [32]. (As discussed in Section 4.1, this is a temporary approach to support our experimentation, and it is not possible to provide sufficient security with the current extension mechanisms.) The repository is available at https://github.com/HainaLi/horcrux_password_manager (full data from our experiments are too large to host there, but is available to interested researchers on request).

### 3.1 Requirements

The driving motivation for our design is to provide strong security against both client and server vulnerabilities, while providing usability that is similar to current password managers. In particular, our design aims to:

- Minimize exposure of user credentials in both time and the amount of code they are visible to.
- Limit the size of the trusted component on the server to a small, auditable core.
- Provide resistance against server compromises by ensuring that even a full compromise of a single password store does not enable at attacker to conduct offline attacks on the user's master password.
- Provide functionality and usability similar to current password managers, including supporting single click logins for sites with stored credentials.

Next, we describe how the password stores are implemented. Section 3.3 describes the client.

### 3.2 Password Stores

For server-side storage of domain credentials, our goal is to ensure that an attacker who obtains a full copy of a single server's store cannot execute a successful attack to learn the user's credentials, or even the domains where the user has stored passwords. Thus, we need to store credentials at the server in a way that an off-line guessing attack is not possible. There are a small set of domains that a user is likely to have credentials for, and an attacker can probably guess domains like facebook.com and gmail.com that will be used by most users. Hence, it is essential that the passwords are stored in a way that does not enable an attacker with access to the store to determine if a guessed master password is correct.

Designing loss-resistant password vaults with such goals in mind has been studied notably by Bojinov et al. [10], who fabricated decoy password sets and by Juels and Ristenpart [25], who introduced the concept of honey encryption to yield plausible-looking but bogus plaintexts for every guess of the encryption key. These designs, however, have been constructed with specific data patterns (i.e. passwords or RSA keys) in mind, and are susceptible to attacks exploiting the differences in the generated distribution of passwords, which is a problem with all static Natural Language Encoders (NLE) [21].

In our case, a better solution is to take advantage of secret sharing since this means all of the actual entries are indistinguishable from randomness. All we need to hide is the locations of actual entries. Our solution is to adopt *cuckoo hashing* [40]. Although cuckoo hashing was not originally designed to support privacy, it can be adapted to meet our goals and provides a solution that is cost-effective for both storage and bandwidth.

In the original cuckoo hashing, a given data point, $x_i$, gets two possible positions, $h_1(x_i)$ in $T_1$ and $h_2(x_i)$ in $T_2$. The hash functions, $h_1$ and $h_2$, are assumed to behave like independent and random oracles for the possible keys in the table. During an insert operation, one empty position is chosen at random to store $x_i$. If both positions are occupied, it follows the cuckoo bird's approach of replacing the current resident with the new resident and repeating and bouncing between the two tables until an empty position is found or when the maximum number of iterations is reached. Pagh and Rodler set this maximum number at $C \log t$, where $C$ is a constant and $t$ is the number of data points. At this point, the entire table is rehashed by choosing a different $h_1$ and $h_2$. Multiple rehashing attempts may be necessary before all the data points find a spot. Successful cuckoo hashing achieves worst case constant lookup and deletion time and amortized constant time for insertions.

In our server design, we use a variant of cuckoo hashing that puts all data in a single table of size $p$, instead of using multiple independent tables. We use $n$ independent hash functions capable of mapping $x_i$ uniformly and randomly into any of the $p$ locations in the table. (Discussion on how parameters are selected is deferred to the end of this subsection.) For each entry, with cuckoo hashing the table can store at any one of (up to) $n$ possible locations:

$$h_1(d, e) = H(\text{str}(1) \;||\; e \;||\; H(d)) \mod p$$
$$\vdots \qquad (1)$$
$$h_n(d, e) = H(\text{str}(n) \;||\; e \;||\; H(d)) \mod p$$

where $d$ is the domain, $e = \text{PBKDF2}(master\_password)$ is the encryption key, $p$ is the table size, and $H$ is a cryptographic hash function (our implementation uses SHA-256).

**Account Entries.** For each account, we store a value in JSON format containing shares of the id tag, username, username length, username tag, password, password length, password tag, and IV.

The tags and IV are for AES-GCM encryption and decryption of the username and password. (The IV is incremented by 10 when encrypting the password.) The data stored in the keystores are not the encrypted plaintext credentials, but a secret share for each of the $s$ keystores, so no information is disclosed in these values. These are padded to a fixed length prior to encryption so no information is disclosed. The username and password fields are padded up to a maximum length (our prototype uses a default maximum of 64 characters), and the length parameters indicate the real length of the credentials. The id tag is just the hash of the domain, $id\_tag = H(d)$, with SHA-256 this is a 256-bit output. An id tag indicating an empty table key location is 256 bits of zeros.

**Initial Table Setup.** The first time the keystores are used, we prepopulate the all the tables with $p$ table key-value pairs that are meant to mask the rows holding real accounts after Horcrux is used. The pre-secret shared values for all the table values are hexadecimal strings of zeros of the length prior to secret sharing. We rely on secret sharing to produce the different shares stored in $p$ rows of the keystores. Using Amazon AWS DynamoDB API's `batchWriteItem` method, we could store up to 25 key-value pairs at a time. All users will start with a table that appears to be full of account entries, and there is no way to distinguish real from empty entries from a single keystore.

**Item Lookup.** To lookup a domain's account information, we simply calculate the table keys associated with the domain using Equation 1 and obtain $n$ table keys. Next, we query each of the $s$ keystores and obtain $n$ values, $v_{s,n}$, from each keystore. After combining the shares, we acquire $n$ values for each of the entries in the table (Note that only the username and passwords were encrypted and the rest do not need to be decrypted):

$$v_1 = dec(\text{combineShares}(v_{1,1}, ..., v_{s,1}), e, IV_1, tag_1)$$
$$\vdots$$
$$v_n = dec(\text{combineShares}(v_{1,n}, e, ..., v_{s,n}), e, IV_n, tag_n)$$

We then compare each of the combined id tags with the hash of the domain, picking the value containing the correct id tag.

**Item Insertion.** In our variant of cuckoo hashing, we have $n$ possible places for the credentials to a domain, $d$. We use cuckoo hashing's collision resolution technique, as described in Section 3.2, to find a "nest" for every credential by calculating the locations for the booted values from its id tag. Inserting new item is the main concern for user-perceived latency, which is discussed in Section 5. Deleting or updating an account, is simply a matter or writing shares of the new value (all zeros for deletion) into the stored location.

**Parameters.** We select the default parameters for our cuckoo hash table for a reasonable balance of storage costs and minimal probability of collisions. Each location in the table can only hold one item. With $n = 2$, the effective load factor is less than 0.5. At higher load factors, cuckoo hashing performance drastically decreases and frequently needs rehashing. With $n \geq 3$, the load factor increases substantially. The value $c^*$ is used to represent the load factor for which the probability that all $p \cdot c^*$ items can be placed in the table. Fountoulakis and Panagiotou proved that for $n = 5$, $c_5^* = 0.992$ [16], meaning that nearly the entire table can be filled before rehashing is needed. For performance reasons, we are also concerned with the expected number of collision resolutions needed. Fountoulakis et al. analyzed the number of insertions needed for a cuckoo hash table, and proved a polylogarithmic bound on the number needed holds for all but negligible probability [17]. For our implementation, the key stores support batch read and right requests (up to 25 elements at a time for AWS' DynamoDB API), so the network cost of increasing the number of hash functions is low, and the local computation cost is also fairly low. Hence, we select $n = 5$.

A 2007 study [15] on the number of accounts owned by the typical user found that the average user as 6.5 passwords, which are shared between 3.9 different sites, and that each user has about 25 accounts. We reason that in 2017, the typical user has more accounts. As a result, we design Horcrux to support a default maximum capacity of 10,000 accounts, and use $p = 10079$ (the first prime number above 10,000), and $n = 5$. This puts our expected maximum random-walk insertion time at around 8 reassignments for a fully-loaded table. Frieze et al. [18] discussed that while breadth-first search gives constant *expected* time, it cannot guarantee sub-polynomial runtime for the insertion of each element. The amortized time complexities are unsuitable for applications that rely on cuckoo hashing to guarantee fast individual insertions. Therefore, we stick to the original cuckoo hashing random-walk insertion algorithm.

### 3.3 Client

The overriding designs goal of the client is to minimize exposure of user credentials to a small, auditable component. Hence, the client is divided into two components: a trusted *core* that needs access to the master password and sensitive credentials, and an untrusted *UI component* that manages interactions with the webpage but cannot access any Firefox API. The core and UI components are completely isolated from each other and can only communicate through the message-passing. For all sensitive computation, The core component creates a NodeJS subprocess which uses its `crypto` library for all cryptographic algorithms[39], as well as AWS and Azure APIs for making requests to keystores.

Next, we describe how to setup the client, and what happens when it is used in the brower to process a login request. Figure 1 illustrates the interactions between the user, browser add-on, and keystores.

**Setup.** To use Horcrux users need to set up servers that store shares of the users passwords. Our prototype add-on prompts users to create accounts on AWS or Azure and create access keys with permissions to use their noSQL keystores (AWS DynamoDB or Azure Table Storage) in a specific datacenter region. More paranoid users will configure keystores themselves using trusted services and spread across jurisdictions. A user may also use their own servers as a keystore, any server that supports the appropriate keystore APIs can be used. Each keystore in initialized as a cuckoo hash table full with shares of empty values as described in Section 3.2.

The user is prompted to provide credentials obtained from AWS or Azure (their *accessKeyID* and *secretAccessKey*) for each keystore, and to create a master password. The core component derives a keystore authentication key from the master password (*MP*) using a Password Based Key Derivation Function (PBKDF2) over 100,000 iterations, following similar practices used by 1Password [1]. The keystore server credentials provided are encrypted using *AuthKey* using AES-256 with GCM mode in base64 encoding, and stored along with the IV, tags, and PBKDF2 salt in a config.json file on the client's device.[2] The IV is newly generated for each encryption, and incremented by 10 for each access key or secret. This file is not accessible to webpages in the browser due to Firefox's isolation policies [34]. Although our design attempts to limit the effectiveness of guessing attacks on the master password, it is still important that users select a strong enough master password to resist at least on-line guessing attacks. Access credentials are randomly generated by AWS or Azure and their base64 encoding prevents them from being susceptible to such dictionary attacks. On-line attacks are limited by the cloud servers limits on incorrect login attempts.

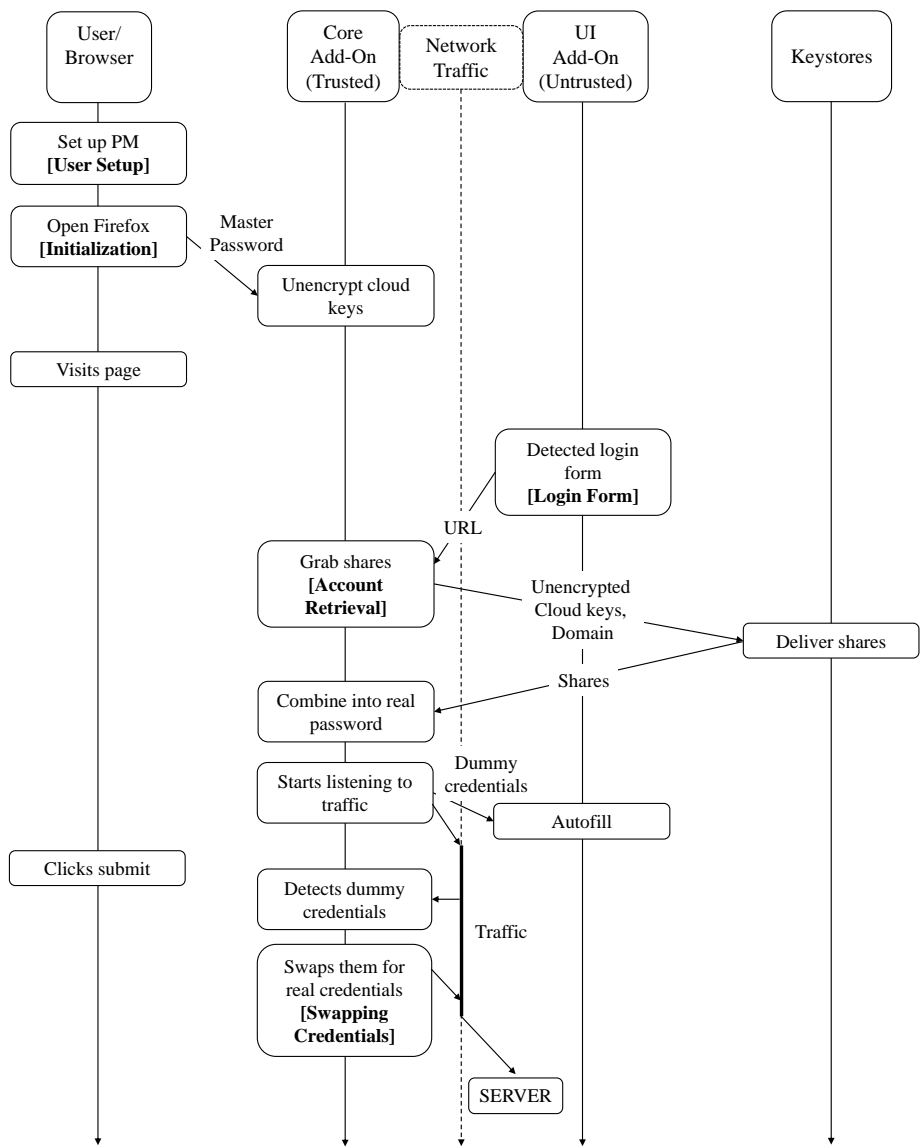

Figure 1: Horcrux protocol

**Initialization.** When a user opens Firefox and begins the browser session, the core component checks for config.json, a file in the add-on containing the user's encrypted keystore credentials. If the encrypted credentials are not present, the setup steps are performed.

---

[2]For AWS keystores, the region in which each set of credentials is configured to use is stored unencrypted. Storing region information is necessary in order to use the AWS API, and encrypting the region name would present an opportunity for an adversary to perform an offline dictionary attack on the master password. We view the risk of leaking the keystore region to an adversary who acquires the encrypted configuration file as much less serious than any risk of enabling an off-line attack on the master password.

Otherwise, the core component prompts the user for their master password. The encrypted credentials are read from `config.json` and stored as strings in the core component's private memory. The master password is used to derive the 512-bit *AuthKey* using `PBKDF2` over 100,000 iterations. The *AuthKey* is then used as the decryption key to an `AES-256` cipher, and the core component deciphers each encrypted credential. The *AuthKey* is kept in the core component memory for the duration of the session (it is kept until browser is closed) to avoid needing to repeatedly request the master password.

Next, the core component verifies the user's keystore credentials by initiating a connection to each keystore. If the request fails, the user is presumed to have given an incorrect master password, and the core component returns to prompting for the master password. There is no enforced limit on password attempts at the client side, as all verification is done online and subject to AWS or Azure policies for making many requests with incorrect credentials.

**Login Form and Account Retrieval.** When the browser loads a webpage, the UI component finds login forms on the page by searching for forms with an input `type=password` and uses heuristics to exclude registration forms based on the form title and input labels. Once a login form is found, then the UI component notifies the core component with the URL of the containing page. The core component proceeds to query to keystores to check for the presence of an account associated with the domain, as previous described in Section 3.2. For the protocol, we assume that each domain only has one account associated with it. See the discussion on supporting multiple usernames per domain in Section 7.

Although it would be simplest and most secure to wait until the user submits the form to start the process of obtaining and reconstructing the account credentials, we considered this unacceptable for user experience. If we fetch the password shares after the user clicks submit, the user would experience the delay of fetching and reconstructing shares as well as the usual server response latency. Further, due to the indeterministic nature of JavaScript, we cannot guarantee that the core component would receive the "user clicked" signal from the UI component before the login traffic leaves the browser.

**Swapping Credentials.** After obtaining credentials for a login form, the core component sends dummy credentials to the UI component to be autofilled in the form, and starts actively listening to network traffic. The UI component autofills the fields in with dummy user and password strings by setting the elements' `value`. These dummy values are different for each user but the same for every website that the user visits. Hence, the actual credentials are never exposed to the DOM, but a user will receive visual feedback that stored credentials are available for the form. The idea of avoiding exposure of passwords to injected scripts by using dummy credentials to autofill forms and replacing them in network traffic was previously suggested by Stock and Johns [46].

The core component waits for the user to click submit, at which point all instances of the dummy username and password in the request traffic are intercepted swapped for the real credentials before TLS encryption. Before inserting the real credentials, the core component checks that the target domain matches the credentials' domain and that the request is secure (the target URL is HTTPS and the dummy username and password are not used as URL parameters in a GET request).

**Enrolling New Accounts.** If the user has not already registered with the domain, Horcrux generates a strong password using random bytes from NodeJS' crypto library (users may override password generation to provide a user-generated password manually). The generated password is prefixed with a fixed set of characters (`AaBa12#$`) which satisfy the majority of web account password creation policies.[3] The credentials are then stored using cuckoo hashing's insert algorithm, as described in Section 3.2.

## 4 SECURITY ANALYSIS

Horcrux is designed to provide strong protection of user credentials against the realistic threat model described in Section 2.2. Here, we analyze how well it resists both client-side (Section 4.1) and server-side (Section 4.2) adversaries.

### 4.1 Client Security

Horcrux's overriding design goals are to minimize attack surface and decentralize trust as much as possible. To achieve this goal, Horcrux is split into two components, a trusted component that has access to the user's entered master password and secrets derived from it and obtained from the keystores using credentials obtained from the master password, and an untrusted component that manages the user interface and everything else that does not involve any sensitive data. The main strategies to minimize the client-side attack surface are to limit the window of opportunity for an attacker to steal the login credentials and to make the client code that has access to the master password as small and simple as possible so that it may be realistically audited. The entire code for the trusted component is around 800 non-comment lines of JavaScript.

**Password Exposure.** Because the password is never inserted into the DOM, it is not vulnerable to reading by injected scripts. Password managers that reveal the real credentials earlier in the process are limited to checking the form action at the time of the autofill, so could be vulnerable to scripts that dynamically modify the form action after the password has been provided. Horcrux ensures that the password is only ever sent to the correct domain by checking the URL in the POST request in the intercepted network traffic. The request must use HTTPS, so this ensures (assuming the browser verifies HTTPS certificates and implements TLS correctly) that the real credentials will only ever be inserted into an outgoing network request to the intended domain.

Malicious scripts could also steal the credentials through a reflection attack by changing the action or field of the form or by inserting a bogus form on the page. When the modified or malicious form is submitted, the credentials are sent to an attacker-controlled section of the website. The only way to prevent this type of attack would be through requiring the entire URL path to match the enrolled path (and hope that page has no reflection vulnerabilities), not just the domain. Enforcing a specific path may break many websites that dynamically generate their target URLs.[4] In this respect, Horcrux is not more vulnerable than traditional password managers or manual input of credentials by a user. As mentioned in Section 2.2, we consider an attacker with knowledge of an arbitrary reflection vulnerability on the host server out of scope, and do not know of any effective defense against such a powerful attacker.

---

[3] We have not focused on password generation, and appreciate that generating passwords that pass some site's password rules is a challenging (and annoying) problem, and no single password will satisfy all rules. The problem of generating good and permissable passwords is orthogonal to the password management issues that are our primary research focus.

**Resisting Guessing Attacks on Master Password.** An adversary who compromises the client's host may be able to obtain the config.json file that contains keystore credentials encrypted with a key derived from the user's master password. This file may also be exposed when a user moves it to setup a new device. As discussed in Section 3.3, Horcrux derives an encryption key from the master password using a password-based key Derivation Function (PBKDF2) over 100,000 iterations. This limits the number of guessing attempts an adversary could make, but a sufficiently resourceful adversary could still be able to execute an effective dictionary attack against a weak master password. Since the user-selected master password may be weak, it is important that there is no way for an adversary to do an off-line dictionary attack against it. There should be no way for the adversary to check if a guessed master password is correct without sending a request to an external service such as an attempt to authenticate with a keystore server. We ensure this by being careful to never encrypt anything with keys derived from the master password that is distinguishable from randomness to an off-line adversary. An adversary can only verify their guess on the master password by deriving the key, using it to decrypt access credentials, and performing an online query with the credentials. The access credentials are random bit strings, so an adversary cannot determine if a guess on the master password is successful without submitting those credentials to a guessed keystore server. This is slow and expensive, and also subject to methods the keystore server should use to limit the number of inauthentic requests attempted (i.e., "throttling"). Neither Amazon nor Microsoft makes these mechanisms public, but at worst, they are no more than the limits for regular requests. AWS limits the steady-state request rates to 1000 requests per second using the token bucket algorithm [5]; Azure limits read requests to 15,000 per hour and write requests to 1,200 per hour [31].

**Security Limitations to a Firefox Add-on.** Browsers add-ons naturally limit the attack surface by isolating the higher privileged core component from the UI component [7]. Because the UI component runs in an insecure environment, the web pages, they are exposed to more threats despite the fact that each script running on the page can only access its own variables. Cross-extension attacks [11] exploit add-ons that have not properly defined its namespace. When a malicious add-ons is using the same JavaScript namespace of a benign add-ons, it can access and modify all of the benign add-on's global variables, functions, and objects. Recently,

LastPass reported that it had updated its non-mobile browser extensions to fix this vulnerability, which allowed a clever attacker to force the LastPass extension to reveal stored user data [28]. We are aware of the threat that cross-extension poses to Horcrux and avoid using global variables, functions, and objects.

As an add-on subjected to the limitations of Firefox's API, our prototype Horcrux implementation intercepts and changes the network traffic using the same methods available to other add-ons. Since Firefox does not have a specific resolution for conflicts between different add-ons, it is not determined which add-on will have the last opportunity to view and edit the request. Another add-on with traffic interception capabilities that executes after Horcrux, would be able to read the final version of the POST request which includes the real credentials. This is serious risk, but mitigated by the user interactions needed to install an add-on. In February 2015, Mozilla made it more difficult for attackers to run malicious add-ons by only allowing reviewed and signed add-ons to run on browsers [35]. This security feature, however, could be turned off by the user, and an insufficiently paranoid user could be tricked into installing a malicious add-on. Section 7 discusses ways Horcrux could be deployed.

### 4.2 Server Security

Horcrux is designed to avoid trusting any single keystore. Here we consider the risks if a single keystore is compromised, and if all of the user's keystores are compromised.

**Single Keystore Compromise.** If an adversary obtains the complete keystore database from one of the keystores (or several of the keystores, up to the secret sharing threshold), they learn no semantic information other than its size if the user's keystore has expanded beyond the default size. The secret-sharing mechanism means that a single share provides no semantic information. Up to the threshold limit number of keystores may be fully compromised without any loss of credentials. Because of the way we use cuckoo hashing and fill all entries of the table with shares which will appear indistinguishable to an adversary, there is no way for an adversary to tell with entries in the table correspond to real accounts. This prevents a guessing attack on master password with popular domains, since all possible guesses lead to an indistinguishable set of possible locations.

Our design does not hide the access pattern, however. This means an adversary who can observe unencrypted requests to the keystore over time (e.g., the keystore operator itself, or an adversary that compromises a keystore server without detection and maintains a monitor there) would be able to learn common patterns of requests. Such an adversary could perform a guessing attack on the master password with common domains, to look for sets of locations that match the requests. Hiding this access pattern would either require giving up on domain privacy (so the domain is no longer encrypted with the master password, but revealed in cleartext in the keystore request), or using expensive methods such as Oblivious RAM [20] to hide access patterns.

The secret sharing schemes we use are malleable, so an adversary who has complete access to a user's keystore could modify their passwords and credentials (with XOR-secret-sharing this is simple bit flipping; with Shamir-sharing it is more difficult to do in a predictable way). This could prevent users from being able to obtain their passwords, but is not a threat to confidentiality.

---
[4]One mitigation for this that may be compatible with more sites, suggested by Ben Stock, is to verify the field names in the transmission before exposing the user's credentials. This would prevent reflection attacks that exploit pages such as a search page which will only reflect content in certain parameters that do not include the username and password fields, but not help with arbitrary reflection pages.

**Multiple Keystore Compromise.** If all the user's keystores are fully compromised, the adversary can reconstruct the users' encrypted passwords. This seems unlikely to happen from vulnerabilities or insider threats if the user has keystores operated by different cloud services, but may be a risk under subpoena threats (users paranoid about NSLs or subpoenas will want to choose the jurisdictions of their keystores accordingly). Each password is encrypted with the master-password derived key before sharing it; the adversary would still need to do a dictionary attack on the master password to obtain the user's password.

## 5 PERFORMANCE

In this section, we evaluate the performance of Horcrux, focusing on the latency that a user may experience with Horcrux' multi-keystore and secret sharing design.

**Experiment Setup.** We conduct both our microbenchmarks (timings for DynamoDB requests) and actual client latency on an EC2 c4.xlarge node in the Northern Virginia region to simulate user experience. The c4.xlarge nodes are equipped with 4 vCPUs, 7.5 GB of memory [3, 4]. For the microbenchmark tests in Table 1, we send individual read and write requests to keystores located in Northern Virginia, Oregon, Ireland, and Singapore. This models a user paranoid enough to want four keystores spread across multiple jurisdictions to resist state-level attacks. In reality, a paranoid user would also want to use different service providers to host the keystores, but for simplicity of our experiments we use AWS for all of them. Since the main issue is latency to the keystores, it should not have a significant impact on the results if the keystores were hosted by different providers, so long as they provide a similar batch request API as the one we use. Both the store and write experiments are taken sending an individual item to the keystores. For the amount of data that we send, a batch request of up to 25 items (the maximum allowed by DynamoDB) does not take noticably longer than an individual item request.

For the user experience tests (Table 2), we report the latency a user would experience when trying to retrieve credentials and enroll a new account. For credentials retrieval, the timer starts when a form is detected and ends when the password is reconstructed and Horcrux is listening for the submit click. This doesn't include the time required to detect the form, but that time mostly depends on the page load time which depends on the individual host. If the credentials are ready by the time the user clicks submit, then Horcrux will have little impact on end users. If not, the submit request would be delayed until the credentials are ready, and the delay may be noticeable and annoying to users.

For enrolling an account, the timer starts when the user indicates that she wants to store account information and ends when the server responds with "write successful". This corresponds to the point when the credentials would be submitted to the account's host server. Cuckoo hashing inserts can sometimes take multiple read and write round trips to the keystores to find a placement. Here, we assume that each store succeeds on the first try. As discussed in

|           | Virginia    | Oregon      | Ireland     | Singapore   |
|-----------|-------------|-------------|-------------|-------------|
| Write (ms)| 64.3 (5.03) | 409 (21.3)  | 361 (25.1)  | 973 (45.2)  |
| Read (ms) | 65.0 (5.16) | 309 (23.7)  | 356 (20.2)  | 927 (37.0)  |

**Table 1: Time in milliseconds for a write or read request from a node in US East (Northern Virginia).** Results are averages over 10 requests, standard deviations in parentheses.

|                        | US (3)       | World (4)   |
|------------------------|--------------|-------------|
| Retrieve Credentials (s) | 0.80 (0.034) | 1.59 (0.49) |
| Enroll Account (s)     | 1.16 (0.04)  | 2.49 (0.17) |

**Table 2: Time in seconds that an user would experience.** The US column is for a user with three keystores, 2 in Northern Virginia and 1 in Oregon, The World column is for the 4 keystores listed in Table 1). Results are averages over 10 requests, standard deviations in parentheses.

Section 3.2, the system parameters are set so it should be very rare for multiple attempts to be needed.

**Results.** Table 1 presents the results from our microbenchmark latency tests. The duration for each read and write requests depends largely on the distance they are from the client. Since the password cannot be reconstructed until all shares are received (using a paranoid configuration where the secret-sharing threshold is set to the number of shares), the latency experienced by the user will depend on the latency to the furthest keystore.

Table 2 shows the average time in seconds that it takes for Horcrux to retrieve credential for a user with keystores on US coasts and one with globally-distributed keystores in the locations in Table 1. In addition to the retrieval layency, there is significant overhead caused by the Firefox extension and cryptographic computations. Enrolling account information takes almost twice as long as fetching an account. This makes sense because in cuckoo hashing, storing into the table requires two roundtrips to the server. A paranoid user who stores shares of his credentials around the world would have to wait approximately 1.6 seconds after the page loads be able to login. To register a new account with Horcrux, the user needs to wait 2.5 or more seconds, depending on how many spots need to be evicted.

We have not done any user testing yet to know how acceptable these results are (that is, how often does a user click submit to login within the credentials reconstruction time, and how noticeable and annoying is the additional delay when the client has to wait for credentials to be ready). Although we suspect these times are long enough to be noticeable for most users, we note that with the right UI display paranoid users may be willing to wait a second or two to perform a login, and there are many opportunities to improve performance to approach the keystore latencies in Table 1, especially for a deployment that is integrated into a browser rather than running as an add-on as our prototype.

# 6 COMPATIBILITY TESTING

For the credential swapping method to work correctly, the dummy credentials must be visible in the outgoing request traffic. Client scripts could alter the values entered for the username and password in ways that would disrupt recognition. In this section, we describe the SwapScan tool we built to perform compatibility testing (Section 6.1), reports on how we tested Alexa's top million websites (Section 6.2), and present our compatibility results (Section 6.3). SwapScan uses heuristics to find login forms on web sites, and then automates the login process using the dummy username and password. For sites where a test login can be performed, SwapScan checks the request traffic for the dummy username and password strings. We found this test to be successful on 98% of the tested logins, which gives us a high confidence that Horcrux would work on a majority of the web. As a by-product of our compatibility test, we also found many insecure login form development practices, which are not directly relevant to password managers; we report on these findings in Section 6.4.

## 6.1 SwapScan

SwapScan, our login automation tool, is built on top of OpenWPM, a fully-automated, open-source framework for large-scale web scanning [14]. OpenWMP provides scalability by wrapping Selenium instances in a driver that monitors their activity, ensuring that if the web page stalls or if one Selenium crashes, the test would continue. OpenWPM, which was originally intended to measure web privacy from third-party scripts, also includes a proxy and a series of hooks for data collection — two features that were essential for our experiments.

SwapScan starts by using OpenWPM to visit a URL from a list of URLs of the front page of websites, and then uses heuristics to scan the site to attempt to find, fill in, and submit a login form. Our approach is adapted from SSOScan [50], but is more complicated since we are looking for login forms instead of SSO buttons.

**Finding Login Forms.** After visiting the URL, SwapScan attempts to find a login form on the website. The strongest indicator of a potential authentication form is the presence of a HTML element with `type=password` in the form children. Once an authentication form is found, SwapScan determines whether the current form is for login or registration. If there are two password children in the form, then the form is likely a registration form. Other indicators used to separate registration forms from logins include keywords in the title of the page, keywords in the form attributes, and whether the form contains an input with registration topics (e.g., birthdate, security question). SwapScan determines the topic of an input box by matching regexes with element attribute values. If the candidate form is not a login, SwapScan will try other forms on the page.

If no suitable form is found on the current page, SwapScan uses heuristics to click on buttons likely to reveal forms until it either finds one or reaches the maximum number of attempts allowed. Candidate buttons are identified on the page and ranked by matching regular expressions such as `[Ll][Oo][Gg][IiOo][Nn]` with each of the attribute values or the innerHTML of a *visible* node [50]. The likely candidates are ranked by the frequency of the matched regular expressions and the attribute the keywords are found in (e.g., the `innerText` of an element would be a better indicator of the purpose of the login). Invisible elements are not considered because a human user would not interact with them, and single-sign-on buttons are ruled out completely. If none of the attempts lead to a login form, SwapScan records that no login was found for this site and concludes this test.

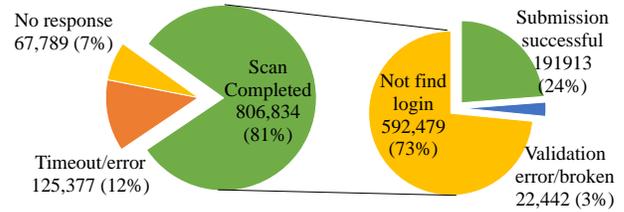

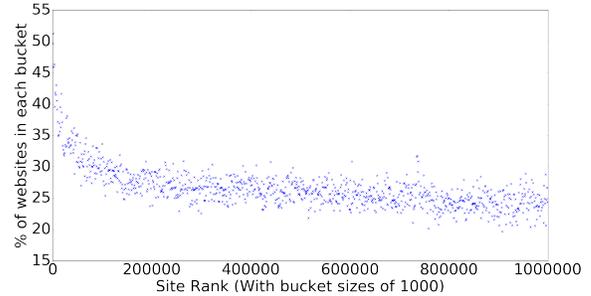

Figure 2: Overview of scan results

Figure 3: **Percent of sites where login is found by popularity rank.** Each bucket is for 1000 sites for which the scan completed.

**Testing Submission.** If a login form is found, it is autofilled with the dummy username and password and submitted. Outgoing traffic is observer to look for the dummy username and password. SwapScan also saves all outgoing traffic from the browser for analysis explained in Section 6.2.

## 6.2 Scanning

In February 2017, we used SwapScan to scan Alexa's top million websites. This took approximately 16 hours using 200 Amazon AWS c4.2large instances, each running 10 parallel scans, averaging around 2 minutes per website.

Figures 2 and 3 summarize the results of the scan. The key finding, discussed in Section 6.3, is that the password swapping approach works on the vast majority of websites (98% in our study). Here, we explain why some websites were not tested and the success rate for finding login forms among those that were tested.

**Nonresponsive Websites.** Out of the 1 million sites attempted, 932,211 (93%) responded with 2xx or 3xx response codes. This result is consistent with OpenWPM's scan result in January 2016, which found that 917,261 (92%) of the websites successfully loaded [14]

The scanner imposes a four-minute timeout for the total time to scan each website. Time outs and errors are affected by network speed and slow server response times (this primarily impacts websites hosted in other countries). SwapScan considers the scan incomplete if it has not completed the full scan itinerary at a maximum depth before the allotted four minutes expire. Errors and timeouts excluded 123,481 (12%) of the websites out from the scan.

**No Discovered Login Forms.** Among the remaining 806,834 sites, SwapScan found 214,355 (27%) websites with native login forms. Some sites do not provide any native login form, either because they do not have user accounts or only support single-sign-on authentication, which we do not include since there is no password to manage. There are several possible reasons, however, why SwapScan could fail to find a login form on a site that has one. The scanning heuristics assume login forms always include an element with type=password, but this does not hold for all login forms. For example, some sites use a two-step login process where the username is collected by the first form, and the password input is only revealed after the visitor submits a username. Our heuristics only consider English-language keywords, so may miss login forms labeled with other languages. Finally, we assume that an important login form would be accessible within a few clicks from the front page of a website.

Figure 3 shows the fraction of websites for which a login was found as the popularity of the site decreases. The success rate reflects the likelihood that more popular sites are more likely to provide user accounts, less likely to rely fully on single-sign-on authentication, and perhaps more likely to be designed in a way that makes the login form easier for an automated scanner to find. We are not able to distinguish among these (and other) possible reasons in our study, however.

Out of the 214,355 sites found with native logins, 22,442 (10%) sites' login did not respond with an outbound request upon submission. This may be due to broken logins or client-side validation errors. The remaining 191,913 sites (90%) were tested for compatibility with Horcrux, and these are the sites analyzed in Section 6.3.

**Comparison with Previous Results.** The largest previous scan of website login forms was Acker et al.'s 2017 study of the top 100,000 websites [2]. They reported finding 48,547 logins on those sites, where our scan only found 25,449. There are several reasons for the difference. We used heuristics to eliminate registration forms, ruling out forms with any indications of a registration-related input. Their scan counted any form with a type=password field as a login form, so counted some forms that were not considered login forms by our scan. Acker et al. counted 2,760 as not reachable within the 51% failed websites, while we considered 16,584 websites as not having completed the scan because they were either broken or did not respond fast enough to make our 4 minute threshold. Our scan required more interactions with the websites because we needed to observe the actual submission traffic, not just find the form, so more websites timed out. Further, we relied on the fact that the website was functioning quickly enough and that the login was reachable from the front page, whereas Acker et al.'s use of crawlers and search engines may have revealed login forms that were not directly accessible from the front page, which boosted their rate of finding logins.[5]

---
[5]We considered using search queries for our scan also, but decided against this because of the unavailability of search engine APIs that allow enough queries for the larger scale of our scan.

## 6.3 Compatibility Results

A site is counted as compatible if both the username and password dummy credentials are recognized in any request. SwapScan records all network traffic during a 3-second duration following a form submission. Of the websites where forms were successfully found and submitted, 187,736 (98%) appeared to be compatible. On these sites, both the dummy username and password were observed in the intercepted traffic so the credential substitution approach should work. While we define compatibility broadly, for security reasons our Horcrux password manager only reveals real credentials in HTTPS POST requests to the correct target domain (Section 4.1).

In cases where multiple requests are sent following a login form submission, we consider the request that contained the most information (e.g., we would favor a request containing both username and password over a request only containing the username). When we only see one of the username or the password in a request, we would check other requests from that website to see if the missing credential is in a different request.

We found some websites that were doing client-side encryption on credentials before sending them out to the server. We found 817 (<1%) of websites sending their requests out with a plaintext username and MD5-hashed password. These websites are counted as compatibale because Horcrux can look for the MD5-hashed dummy password in outgoing traffic and replace it with the MD5-hash of the real password.

**Reasons for Incompatibility.** Of the tested sites, 4177 (2%) appear to be incompatible with credential substitution. For 3266 (78%) of these, the dummy username was observed in intercepted traffic but not the dummy password. We manually examined a sample of ten of these sites, and found that nine of them were sending the dummy password transformed by some function other than MD5 hash. The remaining sampled site had an empty password field, which means that the site was either broken or did not send the password because client-side validation failed. The (probably misguided) inclination for sites to do client-side hashing of passwords explains why not observing the dummy password is the most common reason for incompatibility.

We also manually examined a sample of ten sites selected from the 881 incompatible sites where the password was observed but not the username. Three of these sent requests with empty username fields, one with a transformed username (but not a cryptographic hash). For the six remaining sites, we could not determine the reason why the dummy username had not appeared in the traffic.

One threat to validity is searching for dummy credentials that are not unique and long. We choose credentials with a goal of passing any client-side validation checks, so they could not be long random strings. Still, there is little risk of encountering the same string meant for another context, as our checking scope is limited to requests made in response to our form submissions.

SwapScan found that the password swapping would work on the preponderance (98%) of websites in the top million, and could be successfully deployed in a modern browser to work on today's web. If any popular sites are incompatible (i.e. www.sohu.com) , it may be necessary to include special rules for matching custom client-side password transformations. At worst, users can fall back to manual logins for incompatible sites.

## 6.4 Other Results from Scanning Experiment

Although the goal of our study was to learn about the compatibility of password swapping across the web, as a byproduct of our scan we learned some other interesting things about web logins. Figure 4 summarizes these results.

**Protocol.** From an analysis of the URL destination of these login requests, we found that 102,948 (51%) of logins do not submit a request through HTTPS protocol, which exposes the username and password to man-in-the-middle attacks.

**AJAX.** When a login form uses AJAX to grab the username and password from the form, a header named "X-Requested-With" with a value "XMLHTTPRequest" could be seen in the request. SwapScan used this property to identify 15,707 (8%) websites using AJAX requests, which indicates that JavaScript is used to submit the form data. This is a significant portion of websites that are compatible with Horcrux, but potentially incompatible with other secure login mechanisms, such as Secure Filling [45]. Submitting a form through JavaScript also means that the JavaScript may or may not consider the form action; even if a password manager ensure that the action is secure, it may not be actually used.

**Form action.** Traditional password managers check the form action to ensure that the credentials are being submitted to the correct protocol and domain. Out of the login forms that were scanned, 18,628 (10%) percent of the forms did not have an action listed. For these sites, a non-Horcrux password manager would not be able to check the action to ensure that the credentials are headed to the correct domain. Since static analysis of JavaScript is difficult, the only way to ensure the correct destination for credentials is to dynamically check them, which Horcrux does naturally.

| Category | Percentage |
|---|---|
| HTTPS Protocol | 49% |
| HTTP Protocol | 51% |
| GET Request | 2% |
| POST Request | 98% |
| Other Methods | <1% |
| AJAX Login | 9% |
| Non-AJAX Login | 91% |
| Action Present | 90% |
| Action Not Present | 10% |

**Figure 4: Types of logins found by the scan (percentages are out of the number of completed form submissions).**

## 7 DISCUSSION

Our goal in building the Horcrux prototype and conducting our experiments was to demonstrate the feasibility and practicality of a more secure password manager. Here, we discuss a few issues that we did not address in our prototype implementation but that would be important to handle before deployment.

**Browser Integration.** Our prototype was implemented as a Firefox add-on, but Mozilla is planning on deprecate the Firefox add-on SDK and migrate to WebExtension [36] with the goal of supporting cross-browser compatibility. We built our add-on using the soon-to-be legacy SDK because WebExtension does not currently have support for the majority of low-level APIs [38], some of which we depend on to manipulate HTTPS traffic. WebExtension's `webRequest` API only allows less powerful capabilities such as canceling and redirecting a request, modifying request and response headers, and supplying authentication credentials in-flight [37]. Our add-on needed the capacity to change `POST` request content. Google Chrome's `webRequest` API has similar limitations [33]. We recognize that this design choice is motivated by security—the browser vendors do not want to give third-party, untrusted browser add-ons too much power in case permissions are not reviewed carefully by the user or the developer [7]. In this regard, our design is best-suited for incorporating directly into the browser by changing how the default password manager autofills and stores passwords. Integrating Horcrux into a browser would also eliminate possible traffic visibility among multiple add-ons, which was discussed in Section 4.1. Providing support for mobile devices would also be important, but we have not yet considered that.

**Supporting Multiple Accounts per Domain.** The current Horcrux design does not support multiple usernames for one domain. The keystore scheme can be easily modified to support multiple usernames by including the username when calculating the table keys and picking a larger hash table size. In this design, however, Horcrux needs to know the full username prior to making the request to the keystores. The user would either need to enter the username manually, or Horcrux would need to maintain a local copy of domain and usernames, which defeats the goal of domain and username privacy. An alternative, would be to assume for most domains the user only has one username, but to store something special in the value entry for multi-user domains so the first request would return a list of the usernames instead of the password. Then, those names could be presented to the user to select from in the manager's interface. We believe multiple usernames could be handled without any security compromises.

**Setting up New Devices.** As discussed in Section 3.3, there are two ways to set up new devices to work with Horcrux securely transfer `config.json`, the file containing encrypted keystore credentials, to the new device or re-enter the database access stores into Horcrux. Neither options provide the convenience supported by commercial password managers because the user is not trusting a password manager to store his access keys. Upon entering their master password on the new device, the core component will use `PBKDF2` to derive the key needed to successfully decrypt the keystore credentials.

## 8 RELATED WORK

In this section, we discuss related work on secret sharing and data protection and authentication.

**Secret Sharing and Data Protection.** Although splitting secrets using XOR was known to ancient cryptographers, the first efficient threshold secret-sharing schemes were discovered separately by Shamir [43] and Blakeley [9] in the late 1970s. Our implementation uses Shamir's scheme.

Several prior works have used secret-sharing to protect data. Password-protected secret sharing (PPSS) [6] presents a Public Key Infrastructure (PKI) model for distributing the trust of sensitive data to $t + 1$ hosts, with the ability to protect user data from being reconstructed by at most $t$ compromised hosts. PPSS mentions the utility of distributing user credentials in a password manager setting, though it does not investigate the application of its protocol in relation to common password manager database schemes, such as those found in Gasti et al. [19].

**Password Manager Storage.** Gasti notes the storage weaknesses of numerous password managers such as Chrome's unencrypted local SQLite database, 1Password's credential storage with AES-128 and CBC mode encryption, and KeePass' header-hashing data integrity checks [19]. The paper finds that many popular password managers are poorly suited against IND-CDBA (eavesdropping) or MAL- CDBA (data manipulation) schemes. Gasti categorizes the managers as (1) "those that can be assumed safe on an insecure storage medium" (2) "those that can be used if the underlying storage mechanism provides integrity and data authenticity" and (3) "those that can be used securely only if the underlying storage provides integrity, authenticity and secrecy" . The paper concludes that many password managers used database storage schemes that were vulnerable to read-only and read-write attacks.

**Protecting Passwords from Scripts.** Stock and Johns considered the problem of injected scripts stealing autofilled credentials from password forms, and proposed a client-side defense similar to the password swapping method used by Horcrux [46]. They implemented a prototype Firefox extension to perform password swapping, similar to what is done by our implementation, and conducted an evaluation of the top 4000 websites to determine how many included scripts that accessed password data, and manually inspected some of those scripts to understand whether password swapping was likely to work on those sites. Since their study required manual analysis, it could not scale to a large number of websites. Their implementation used local storage of passwords, and did not consider ways to reduce the size of the trusted client-side component.

## 9 CONCLUSION

Passwords remain essential for web security, and improving the security of password management is an important goal for our community. We have demonstrated that it is possible for a password manager to minimize exposure of passwords to a few hundred lines of simple code, while using secret-sharing to mitigate the threat of server compromises.

## ACKNOWLEDGMENTS

The authors thank Samuel Havron for coming up with the Horcrux name and contributions to the design, implementation, and presentation. This work was partially funded by an award from the National Science Foundation (1422332) and gifts from Amazon, Google, and Microsoft.